%% file: cmb2_pap.tex
\documentclass[onecolumn]{mn2e}  
\usepackage[dvips]{graphics}
\usepackage{epsfig}
\usepackage{epsf}
\usepackage[figuresleft]{rotating}
\begin{document}
%
\input{mymacros.tex}

%
%
%
\newcommand{\leftb}{<\!\!} \newcommand{\rightb}{\!\!>}
\newcommand{\oversim}[2]{\protect{\mbox{\lower0.5ex\vbox{%
  \baselineskip=0pt\lineskip=0.2ex
  \ialign{$\mathsurround=0pt #1\hfil##\hfil$\crcr#2\crcr\sim\crcr}}}}} 
\newcommand{\simgreat}{\mbox{$\,\mathrel{\mathpalette\oversim>}\,$}} 
\newcommand{\simless} {\mbox{$\,\mathrel{\mathpalette\oversim<}\,$}} 
%
%
%
%
\title{On the mass function of star clusters}  
   \author[Kroupa \& Boily]{P. Kroupa$^1$ \& 
C.~M. Boily$^{2}$ 
\\          
$^{1}$Institut f\"ur Theoretische Physik und Astrophysik der
Universit\"at Kiel, D-24098 Kiel, Germany 
\\ 
$^{2}$Observatoire astronomique de Strasbourg, 11 rue de l'universit\'e, 67000 Strasbourg, France}

\maketitle

\begin{abstract} 
Clusters that form in total $10^3\simless N \simless 10^5$ stars
(type~II clusters) lose their gas within a dynamical time as a result
of the photo-ionising flux from O~stars.  Sparser (type~I) clusters
get rid of their residual gas on a timescale longer or comparable to
the nominal crossing time and thus evolve approximately adiabatically.
This is also true for massive embedded clusters (type~III) for which
the velocity dispersion is larger than the sound speed of the ionised
gas. On expelling their residual gas, type~I and~III clusters are
therefore expected to lose a smaller fraction of their stellar
component than type~II clusters.  We outline the effect this has on
the transformation of the mass function of embedded clusters (ECMF),
which is directly related to the mass function of star-cluster-forming
molecular cloud cores, to the ``initial'' MF of bound gas-free star
clusters (ICMF). The resulting ICMF has, for a featureless power-law
ECMF, a turnover near $10^{4.5}\,M_\odot$ and a peak near
$10^3\,M_\odot$. The peak lies around the initial masses of the Hyades,
Praesepe and Pleiades clusters.  We also find that the entire Galactic
population~II stellar spheroid can be generated if star formation
proceeded via embedded clusters distributed like a power-law MF with
exponent $0.9\simless \beta \simless 2.6$.

{\keywords stellar dynamics -- stars: formation -- Galaxy: formation --
globular clusters: general -- open clusters and associations: general
-- early universe}

\end{abstract}

\section{Introduction}
\label{sec:intro}

Star clusters form embedded in gas, and because the star formation
efficiency (SFE), $\epsilon$, is low, they lose a large fraction of
their stars when the gas is expelled. Open and globular clusters thus
contain only a fraction of the original population of stars in the
cluster.

By comparing the binding energy of the embedded cluster with the
binding energy of the post-gas expulsion cluster, Hills (1980) and
Mathieu (1983) showed that if more than 50~per cent of the total mass
($\epsilon<0.5$) is removed rapidly on a time scale shorter than the
crossing time of the stars through the embedded cluster ($\tau_{\rm G}
< t_{\rm cr}$), then the entire cluster dissolves. This argument
assumes the embedded cluster to be close to global virial equilibrium
just prior to gas expulsion, a state that is considered to be the most
realistic (Kroupa 2002a).  The energy argument posed a serious
challenge, because $\epsilon \equiv M_{\rm ecl} / (M_{\rm ecl} +
M_{\rm G}) =0.2-0.4$ by observation (e.g. Lada 1999; N\"urnberger et
al. 2002), where $M_{\rm ecl}$ is the mass in stars that form in the
embedded cluster within a region a few times larger than the
characteristic radius, $R$, of the embedded cluster, while $M_{\rm G}$
is the mass in residual gas within this same region. The problem
arises because bound star clusters like the Pleiades and Hyades
exist. They are sufficiently populous to have contained a few O~stars
that photo-ionise the cloud core rapidly which, as a consequence of
being thus heated to $10^4$~K, expands thermally at the sound speed,
$v_s\approx 10$~km/s, which is larger than the velocities of the stars
in the embedded cluster.

This rather severe problem was re-addressed in a pioneering study by
Lada, Margulis \& Dearborn (1984, hereinafter LMD) using $N-$body
computations of clusters containing up to $N_{\rm ecl}=100$ stars
embedded in a gas potential that was removed over various
time-scales. LMD, and subsequent work with much higher $N_{\rm ecl}$
by Goodwin (1997, 1998) and Geyer \& Burkert (2001), confirmed Hills'
and Mathieu's result.  LMD noted, however, that a small bound core of
stars formed even if gas expulsion was rapid. Their small total number
of stars precluded a more detailed study of these cores. Improved
$N-$body algorithms and computer power allowed this problem to be
attacked anew. Kroupa, Aarseth \& Hurley (2001, hereinafter KAH)
performed computations with $N_{\rm ecl}=10^4$ stars in 100~per cent
primordial binaries and a realistic stellar IMF using a state-of-the
art high-precision $N-$body code and assuming that the mass-dominating
residual gas ($\epsilon=0.33$) is expelled on a thermal expansion
timescale. They showed that Orion-Nebula-like embedded clusters easily
evolve to Pleiades-like open clusters. These contain a fraction
$f_{\rm st}=0.2-0.3$ of the stars originally in the embedded
cluster. The KAH models include stellar evolution and a realistic
Galactic tidal field which unbinds a larger fraction of the expanding
stellar population than in the previous models that neglected the
tidal field contribution. The KAH result is interpreted to be a
consequence of the formation of a stellar core through stellar
encounters that redistribute the energy prior, during and after gas
expulsion (Boily \& Kroupa 2002).

Direct observational evidence for the loss of a substantial fraction
of the stars from an embedded cluster is provided by Elson et
al. (1987), who study a sample of 10 massive young clusters in the LMC
and find that many if not all of them have up to 50~per cent of their
total mass in a halo of unbound stars.  Depending on the distribution
of embedded cluster masses and the gas-removal time-scale, such
expanding populations can add a kinematically hot component to the
field population of a galaxy. This may explain the hitherto not
understood age--velocity dispersion relation for solar-neighbourhood
stars and possibly the origin of the Galactic thick disk (Kroupa
2002a).  The formation of star clusters and the cluster mass function
thus become important for shaping galaxies. The notion that star
clusters form by expelling a large fraction of their stars also calls
into question conclusions about the dominant mode of star formation
based on studies of the number of open and globular clusters and their
evaporation rate through two-body relaxation. Thus, for example, the
Galactic population~II spheroid is typically interpreted to be the
result of a clustered mode of star formation (the globular clusters)
and an isolated mode (the population~II field stars).

The mass function of young star-clusters (CMF) is also fundamentally
important for constraining star-cluster formation theories and for
dynamical population synthesis studies that aim to account for the
binary properties of a galaxy's population (Kroupa 1995a).
Furthermore, it is becoming apparent that star clusters play a
crucially important role for the chemical evolution of galaxies:
long-lived and thus initially massive clusters lead to a significantly
enhanced production rate of type~Ia supernovae through the formation
of short-period double-white-dwarf and giant--white-dwarf systems due
to encounters (Shara \& Hurley 2002). It is thus essential to know the
distribution and number of long-lived clusters.

The CMF is not well constrained though, with log-normal or power-law
forms being discussed for various star-forming galaxies
(e.g. Elmegreen et al. 2000; Fritze-v. Alvensleben 1999; Whitmore et
al. 1999).  Inferring the CMF is very difficult.  Observational
biases, such as obscuration by dust, flux limits, and radial
truncation through the limited survey area together with the unknown
stellar-dynamical state of the cluster make inferences about the mass
highly uncertain.  There are also significant theoretical
uncertainties, such as the rate of star-cluster dissolution through
internal evolution and external agents, and uncertainties in stellar
models that define the mass-to-light ratia of young clusters and their
ages.

The above discussion makes it apparent though that because cluster
birth is associated with a large fraction of stars being lost from
their embedded cluster, the physically relevant distribution function
is the {\it MF of embedded clusters} (ECMF). The ECMF accounts for all
the stars born in a star-cluster. The ECMF differs from the standard,
or classical {\it ``initial'' mass function of bound and gas free star
clusters} (ICMF). The ICMF is estimated either by observing ensembles
of young star clusters with the above mentioned uncertainties, or by
inferring initial cluster masses by correcting observed cluster masses
for mass-loss through stellar and ``standard'' stellar-dynamical
evolution.

This contribution addresses the issue of how the ICMF is related to
the ECMF.  We define the ICMF to be {\it the MF of an ensemble of
gas-free star clusters in which each cluster is bound and virialised
within the tidal field but not older than a critical time $t_{\rm
ICMF}$}.  As the critical time we take the shortest between the
stellar-evolution time-scale, $t_{\rm st}$, and the initial half-mass
relaxation time, $t_{\rm rel}$, of the virialised object.  Restriction
to ages younger than the stellar-evolution time-scale (about
$10^{7-8}$~yr) is necessary because by this time a cluster loses more
than about 10~per cent of its mass through stellar evolution (Kroupa
2002b). For clusters with $t_{\rm st} > t_{\rm rel}$ the restriction
to ages younger than $t_{\rm rel}$ is necessary in order to limit mass
loss from the cluster through two-body relaxation. The definition of
the ICMF is important for the construction of the ICMF from
observational data. As shall be shown in this paper, structure in the
ICMF allows star-cluster formation theories to be constrained.

The article is laid out as follows. In \S~\ref{sec:etoi} the concept
of inferring an {\it initial} cluster mass for an observed present-day
cluster using the {\it classical}, or usually adopted, method is
explained, and in \S~\ref{sec:classes} we consider characteristic
time-scales involved in the formation of star clusters. We use these
to sub-divide the ensemble of embedded clusters into three types. This
sub-division is used in \S~\ref{sec:icmf} as an ansatz for
parametrising the transformation of embedded cluster mass to (initial)
bound cluster mass after removal of the residual gas, allowing the
transformation of the ECMF to the ICMF to be computed.  The practical
construction of the ICMF is outlined in \S~\ref{sec:toicmf}.
\S~\ref{sec:stsph} presents a short application of the concepts raised
with this contribution by addressing the origin of the population~II
stellar spheroid.  The conclusions are presented in \S~\ref{sec:conc}.

\section{The classical initial mass of a cluster}
\label{sec:etoi}
The ``standard'' method of inferring the initial mass of a star
cluster in the solar vicinity amounts to estimating its
stellar-dynamical evolution which is mostly driven by evaporation of
stars from the cluster through two-body relaxation, and by taking into
account mass loss from evolving stars, and a standard (solar-vicinity)
Galactic tidal field (Terlevich 1987; de la Fuente Marcos 1997;
Portegies Zwart et al. 2001).  The presently observed state of a
cluster can thus be mapped to an initial state, which corresponds to
the case $\epsilon=1$ and initial dynamical equilibrium. Tidal
shocking from passages of the cluster through the plane of the
Milky-Way disk, which lead to additional mass loss from the cluster,
can be readily incorporated in those cases where a cluster is known to
have a significant out-of-plane motion.

\begin{figure}
\begin{center}
\rotatebox{0}{\resizebox{0.6 \textwidth}{!}
{\includegraphics{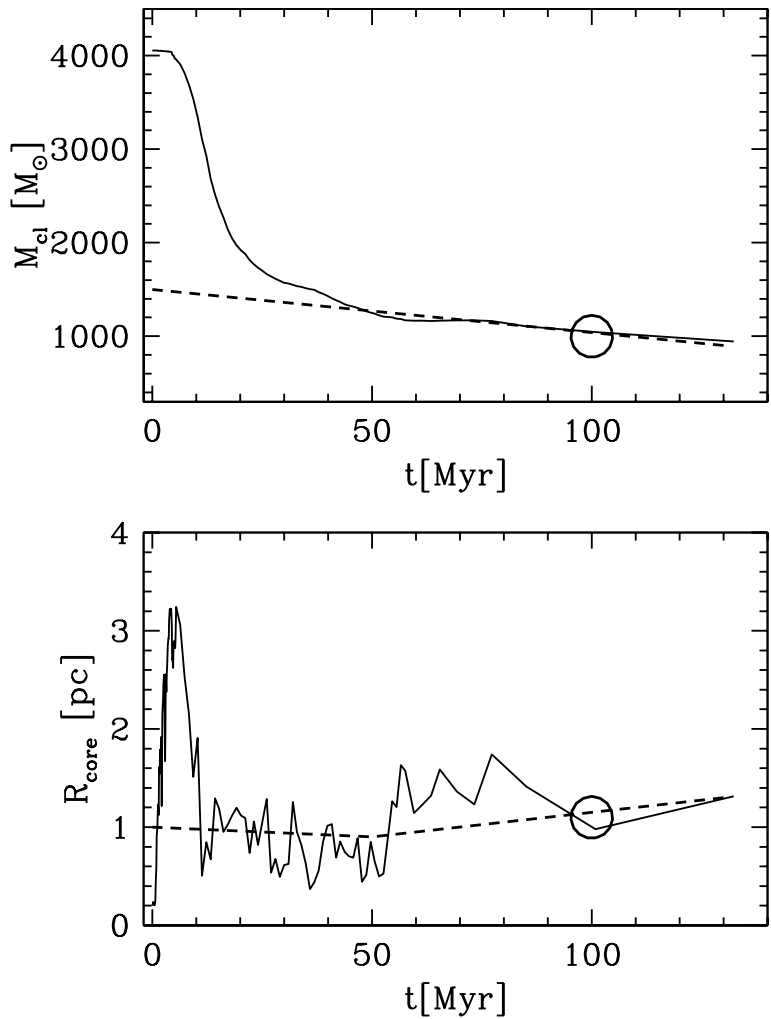}}}
\vskip 0mm
\caption
{The embedded cluster has a significantly different mass and core
radius than the classical initial cluster. The solid line in the upper
panel shows the evolution of the stellar mass in an Orion-Nebula-like
embedded cluster ($M_{\rm ecl}=4054\,M_\odot$, model~A in KAH), while
the solid line in the lower panel shows the evolution of the core
radius. The open circle is a hypothetical open cluster datum.  The
dashed lines indicate the classical approach to estimate the initial
configuration. The classical approach only takes account of mass loss
in a tidal field driven by standard evaporation through two-body
relaxation and stellar evolution, as well as the rare ejected stars.
The classical approach infers an initial mass $M_{\rm
icl}\approx1500\,M_\odot$ and an initial core radius $R_{\rm
core}\approx 1$~pc.  }
\label{fig:clev}
\end{center}
\end{figure}

The contrast between the actual and the classical evolution tracks is
demonstrated in Fig.~\ref{fig:clev}. The figure shows the evolution of
an embedded cluster with $\epsilon=0.33$ that expels its gas faster
than a dynamical time after an embedded period lasting for 0.6~Myr
(model~A in KAH). The model cluster is on a circular orbit at the
solar distance in the plane of the Milky-Way disk.  The mass in bound
stars decays rapidly after gas expulsion, while the core radius
expands significantly. The bound cluster stabilises by about 40~Myr
after core-contraction to $R_{\rm core}\approx1$~pc near a mass of
$1500\,M_\odot$. Further evolution is driven by standard evaporation
through two-body relaxation and stellar evolution which inhibits
further significant core contraction.

Now assume there exists, in the solar vicinity, a star cluster on a
circular orbit in the plane of the disk and with an age of 100~Myr
inferred from its stellar population. It is represented by the open
circles in Fig.~\ref{fig:clev}. The classical evolution tracks would
fit this cluster with an {\it initial} model represented by $M_{\rm
icl}\approx 1500\,M_\odot$ and $R_{\rm core}\approx 1$~pc. A
stellar-dynamical computation beginning from this initial state would
reproduce the cluster. The initial state of a cluster is thus
degenerate as far as the mass and radius are concerned, but additional
information such as the distribution of binary-star periods within the
observed cluster helps to lift some of the degeneracy (Kroupa 2000).
Here the argument is essentially that the widest binaries constrain
the maximum concentration of the cluster.

As an example, Portegies Zwart et al. (2001) infer the {\it initial}
mass of the Pleiades (about 100~Myr old), Hyades (about 600~Myr) and
Praesepe (400-900~Myr) to have been about the same ($1600\,M_\odot$),
but such classical estimates do not account for the gas-expulsion
($\epsilon<0.5$) and associated loss of stars.

Although {\it initial} cluster parameters {\it do not reflect a
physically relevant initial configuration}, they nevertheless provide
a potentially powerful path for probing the physics involved in
star-cluster formation.

\section{Timescales and classes of embedded clusters}
\label{sec:classes}

General consideration of the crossing time-scale and the time-scale
involved in the assembly of star clusters relative to the
gas-expulsion time-scale allows an assessment of the fraction of stars
lost from an embedded cluster.  This approach leads to a rough but
useful sub-division of embedded clusters into three classes.

The introduction of a few generic quantities is useful for the
following discussion.  The nominal global crossing time of an embedded
cluster can be defined as $t_{\rm cr} = 2\,R/\sigma_{\rm 3D}$, where
the nominal global three-dimensional velocity dispersion of the stars
in the embedded cluster is $\sigma_{\rm 3D} = (G\,\left(M_{\rm
ecl}+M_{\rm G}\right) / R)^{1/2}$, and $R$ is the gravitational radius
of the embedded cluster ($G=0.0045$~pc$^3 M_\odot^{-1}$ Myr$^{-2}$).
Perusal of observational data (Lada 1999; Clarke, Bonnell \&
Hillenbrand 2000; Wilking 2001) shows that $R\approx0.5-1.5$~pc
despite a variation of $M_{\rm ecl}$ by many orders of magnitude. We
therefore set $r=1$~pc, and consider $M_{\rm ecl}$ to be the primary
variable.  A constant SFE of $\epsilon=0.33$ is adopted for all
cluster masses.  This appears to be consistent with observational data
for low-mass embedded clusters (Lada 1999) and massive star-burst
clusters (N\"urnberger et al. 2002).  The number of stars in an
embedded cluster is $N_{\rm ecl} = M_{\rm ecl}/m_{\rm av}$, where
$m_{\rm av}=0.4\,M_\odot$ is the average stellar mass.  The
star-cluster formation timescale is $\tau_{\rm sf}^{\rm cl} =$ one to
a few~Myr (Hartmann 2001).  The time over which a significant fraction
of the gas is removed from an embedded cluster, the gas-expulsion time
scale, $\tau_{\rm G}$, depends on the presence of O~stars and on the
depth of the potential well. Details are still very uncertain, but
three general types of behaviour allow a rough sub-division of
embedded clusters:

\renewcommand{\labelitemi}{$\bullet$}
\begin{itemize}

\itemsep=2mm

\item {\bf Type~I:} These sparse embedded clusters contain no O~stars,
and thus $N_{\rm ecl}\simless 10^3$ assuming a standard stellar IMF
(Kroupa 2002b).  Sparse embedded clusters remove their gas on a
time-scale comparable to the cluster-formation time-scale because the
primary mechanism is the accumulation of the increasing number of
outflows as the stellar population builds up (Matzner \& McKee 2000).
This is comparable to the nominal crossing time. We thus have
\begin{equation}
\tau_{\rm G} \approx \tau_{\rm sf}^{\rm cl} \approx t_{\rm cr}.
\end{equation}
Such systems may thus not be mixed and appear sub-structured. These
clusters lie in the region between adiabatic and explosive
evolution. Members of this class are the embedded clusters in Orion
cloud~L1630 (Lada \& Lada 1991) and the embedded cluster $\rho$~Oph
(Bontemps et al. 2001).

In the adiabatic limit and for an isolated cluster all initially bound
stars remain bound despite significant expansion (Hills 1980; Mathieu
1983). A tidal field from the parent molecular cloud and the Milky Way
unbinds a significant fraction of the expanded cluster though, so that
the fraction of stars, $f_{\rm st}$, that remain bound after adiabatic
gas removal will be reduced significantly.  This fraction depends on
the details of the mass profile and the tidal field, and here it is
assumed very roughly that the ``initial'' cluster contains a fraction
$f_{\rm st}\approx0.5$ of $N_{\rm ecl}$.

In the explosive limit and with a solar-neighbourhood tidal field and
stellar evolution, $f_{\rm st}<0.3$ (KAH).

Initial clusters with $N_{\rm icl} = f_{\rm st}\,N_{\rm ecl}\simless
500$ evaporate completely due to two-body relaxation within less than
1~Gyr (Terlevich 1987; Kroupa 1995b, de la Fuente Marcos 1997). Such
clusters would not be evident in star-cluster catalogues because their
low density precludes easy identification above the background
Galactic field density.

\item {\bf Type~II:} $10^3\simless N_{\rm ecl}\simless 10^5$.  These
embedded clusters are rich enough to contain between one and about
150~O~stars but are not so massive as to have a nominal velocity
dispersion larger than the sound-speed of the ionised gas.  The
nominal crossing time is shorter than the cluster-formation time-scale
($t_{\rm cr} < \tau_{\rm sf}^{\rm cl} \approx 1-$~few~Myr). The
cluster is thus approximately in virial equilibrium and mixed for
times $t>t_{\rm cr}$, since most of the proto-stars have enough time
to cross the system and ``virialise'' before the gas is expelled.

Gas expulsion occurs on a time-scale given by the time it takes for
the hot ($10^4$~K) ionised gas to expand outwards as a result of the
overpressure. This expansion occurs, approximately, with the
sound-speed, $v_s\approx10$~km/s
\begin{equation}
\tau_{\rm G} \approx {R \over v_s} \simless t_{\rm cr}.
\end{equation}
The gas is thus removed ``explosively'', implying $f_{\rm st} < 0.3$
(KAH).  

An example of such a system is the ONC (Hillenbrand \& Hartmann 1998).

\item {\bf Type~III:} $N_{\rm ecl}\simgreat 10^5$. These clusters
contain more than a few hundred O~stars and have a velocity dispersion
$\sigma_{\rm 3D} > v_s$.  The nominal velocity dispersion is thus
larger than the sound speed of the hot ionised gas, so that the gas
cannot leave through thermal expansion.  Supernovae may be needed to
do the job. In this case the gas-expulsion time-scale may take
several~Myr after the first supernova explodes, if more than one
supernova is needed (Goodwin, Pearce \& Thomas 2002). Thus
\begin{equation}
\tau_{\rm G} \gg t_{\rm cr}.
\end{equation}
When this is true, then the embedded cluster reacts approximately
adiabatically and $f_{\rm st}\approx0.5$.  An example of a very young
cluster of this type is R136 in the LMC which appears to have already
removed most of its gas (Hunter et al. 1995; Selman et al. 1999).

\end{itemize}

Even if the details are still very uncertain, the above sub-division
demonstrates that a variation of $f_{\rm st}$ with $M_{\rm ecl}$
should be expected. The next section will demonstrate that such a
variation leaves its imprint in the ICMF. This is why the ICMF is of
interest and may be used to infer the function $f_{\rm st}(M_{\rm
ecl})$.

\section{The initial cluster mass function}
\label{sec:icmf}

In what follows two examples for $f_{\rm st}(M_{\rm ecl})$ are
considered to outline the associated transformation from an ECMF to an
ICMF. These examples are intended as a demonstration of the
information carried by structure in the ICMF.

\begin{figure}
\begin{center}
\rotatebox{0}{\resizebox{0.6 \textwidth}{!}
{\includegraphics{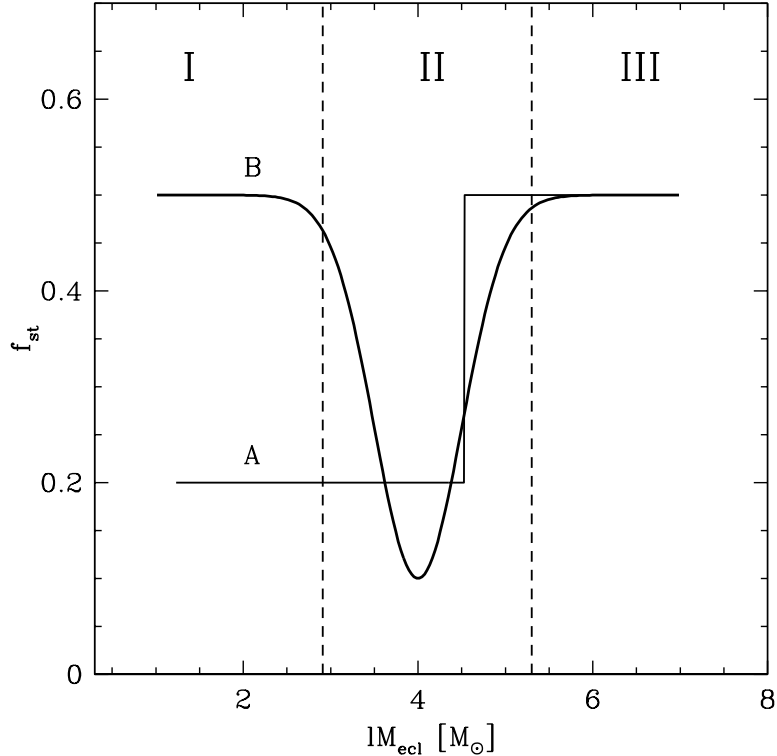}}}
\vskip -20mm
\caption
{
Transformation factor as a function of embedded cluster mass
for the two examples considered in the text. 
}
\label{fig:fst}
\end{center}
\end{figure}

The number of embedded clusters in the mass interval $M_{\rm ecl}, M_{\rm
ecl}+dM_{\rm ecl}$ is
\begin{equation}
dN^{\rm ecl} = \xi^{\rm ecl}(M_{\rm ecl})\,dM_{\rm ecl},
\label{eq:mf}
\end{equation}
where $\xi^{\rm ecl}(M_{\rm ecl})$ is the ECMF. Assuming all embedded
clusters form bound clusters so that $f_{\rm st}(M_{\rm ecl})$ is
continuous, the ICMF follows from
\begin{equation}
\xi^{\rm icl}(M_{\rm icl}) =\xi^{\rm ecl}(M_{\rm ecl}) \, 
                {dM_{\rm ecl} \over dM_{\rm icl}}.
\label{eq:icmf}
\end{equation} 
To fix ideas and based on the discussion in \S~\ref{sec:classes} we
consider the following two examples for the transformation function
\begin{equation}
M_{\rm icl} = f_{\rm st}(M_{\rm ecl})\,M_{\rm ecl},
\label{eq:transf}
\end{equation}
between embedded and initial cluster mass, assuming the average
stellar mass, $m_{\rm av}$, remains unchanged.  The two examples are
plotted in Fig.~\ref{fig:fst}.  Note that the present example
functions $f_{\rm st}(M_{\rm ecl})$ have no physical meaning apart
from providing simple parametrisations of the type of behaviour
expected (\S~\ref{sec:classes}).

\vspace{5mm}

\noindent{\bf Example A:}
\begin{equation}
f_{\rm st} =   \left\{ 
               \begin{array}{l@{\quad:\quad}l@{\quad,\quad}l}
               0.2 & N_{\rm ecl} < 8.5\times10^4 & 
               {\rm explosive\; evolution\;
               for \; types~I\; and~II},\\
               0.5 & N_{\rm ecl} \ge 8.5\times10^4 & {\rm adiabatic\;
               evolution\; for\; type~III, \;
               assuming \; winds \; play\; no\; role}.
               \end{array}\right.
\label{eq:fstA}
\end{equation} 
Note that this example does not allow for near-adiabatic evolution of
sparse clusters.

\vspace{5mm}

\noindent{\bf Example B:}
\begin{equation}
f_{\rm st}(M_{\rm ecl}) = 0.5 - 0.4\,{\cal G}
             (lM_{\rm ecl};l\sigma_{M_{\rm ecl}},lM_{\rm ecl}^{\rm expl}),
\label{eq:fstB}
\end{equation}
where ${\cal G}(lM_{\rm ecl})$ is a Gaussian function in
$lM_{\rm ecl} \equiv {\rm log}_{10}(M_{\rm ecl}/M_\odot)$,
\begin{equation}
{\cal G}(lM_{\rm ecl};l\sigma_{M_{\rm ecl}},lM_{\rm ecl}^{\rm expl}) = 
              e^{ - {1\over2} \left({lM_{\rm ecl}-lM_{\rm ecl}^{\rm expl}
                 \over \sigma_{lM_{\rm ecl}}} \right)^2 },
\label{eqn:gaus}
\end{equation}
with mean $lM_{\rm ecl}^{\rm expl} \equiv {\rm log}_{10}(M_{\rm
ecl}^{\rm expl}/M_\odot)$ and variance $\sigma_{lM_{\rm ecl}}^2$ and
scaled such that ${\cal G}(M_{\rm ecl}=M_{\rm cl}^{\rm expl})=1$.

This form for $f_{\rm st}$ together with $lM_{\rm cl}^{\rm expl}=4$
and $\sigma_{lM_{\rm ecl}}=0.5$, assumes that embedded clusters of
type~I and~III evolve approximately adiabatically as the gas is
removed ($f_{\rm st}\approx 0.5$), while embedded clusters with a mass
near $M_{\rm ecl}^{\rm expl}=10^4\,M_\odot$ suffer explosive gas loss
and consequently lose up to 90~per cent of their stars.

\vspace{5mm}

\begin{figure}
\begin{center}
\rotatebox{0}{\resizebox{0.6 \textwidth}{!}
{\includegraphics{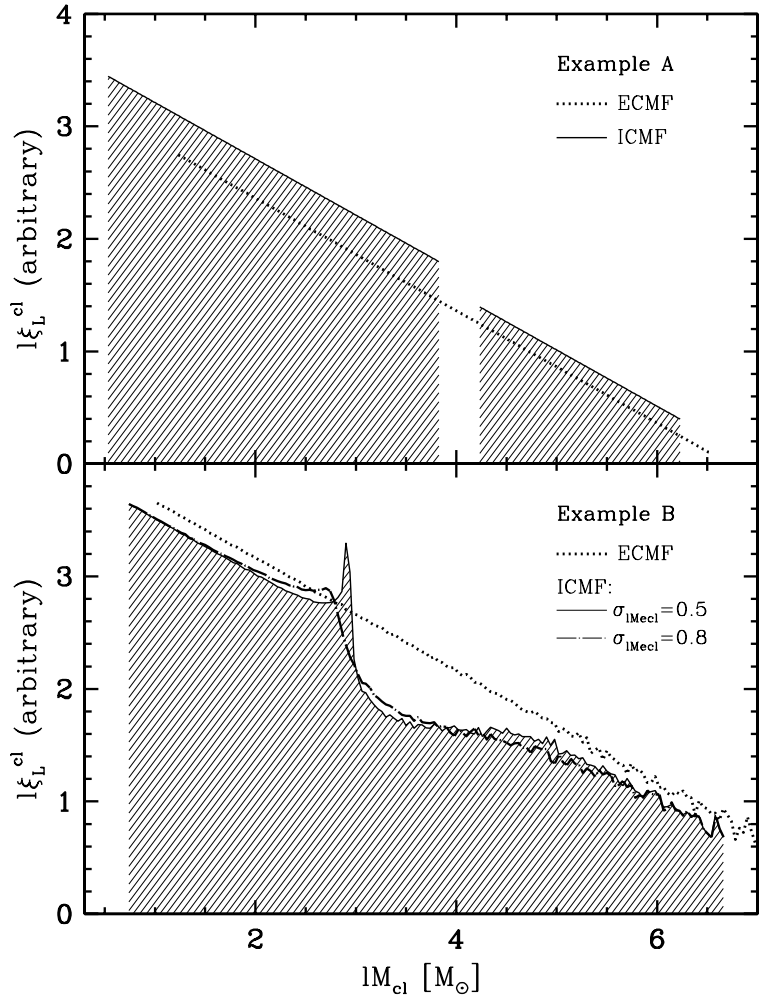}}}
\vskip 0mm
\caption
{The logarithmic embedded cluster mass function (ECMF, $M_{\rm
cl}=M_{\rm ecl}$) and the logarithmic initial cluster MF (ICMF,
$M_{\rm cl}=M_{\rm icl}$), for the two examples considered in the text
[$l\xi^{\rm cl}_{\rm L}(lM_{\rm cl}) \equiv {\rm log}_{10}(\xi^{\rm
cl}(lM_{\rm cl}))$, $\xi^{\rm cl}_{\rm L} = (M_{\rm cl}\,{\rm
ln}10)\,\xi^{\rm cl}$]. {\bf Upper panel:} the ECMF is
eq.~\ref{eq:ecmf} with $\beta=1.5$, and the ICMF is
eq.~\ref{eq:icmf}. {\bf Lower panel:} the ECMF and ICMF are obtained
by picking $10^6$ cluster masses at random from the ECMF
[eq.~\ref{eq:ecmf} with $\beta=1.5$] and transforming each embedded
cluster mass to the initial cluster mass (eq.~\ref{eq:transf}), and
binning both samples to create the histograms shown. The ICMF is
obtained assuming $lM_{\rm ecl}^{\rm expl}=4$ for two values for
$\sigma_{lM_{\rm ecl}}$.  }
\label{fig:mf}
\end{center}
\end{figure}

Fig.~\ref{fig:mf} shows a power-law ECMF,
\begin{equation}
\xi(M_{\rm ecl}) = A\,M_{\rm ecl}^{-\beta},
\label{eq:ecmf}
\end{equation}
and the resulting ICMF. 

In example~A the resulting ICMF has a gap in the mass range $6800$ to
$17000\,M_\odot$, which arises because embedded clusters with masses
$M_{\rm ecl}\ge3.4\times10^4\,M_\odot$ lose 50~per cent of their
stars, while embedded clusters with $M_{\rm ecl}<3.4\times
10^4\,M_\odot$ lose 80~per cent of their stars.

For example~B the ICMF has a depression over a similar mass range but
a peak near $M_{\rm icl}=10^3\,M_\odot$ that can be very sharp if
$\sigma_{lM_{\rm ecl}}\simless 0.5$. These features follow by
differentiating eq.~\ref{eq:transf}, but the Monte-Carlo approach
adopted for the lower panel of Fig.~\ref{fig:mf} can readily be
extended to include more complex effects, such as merging clusters.
The turnover of the ICMF near $10^5\,M_\odot$ is noteworthy
considering empirical evidence for a turnover of the young-cluster MF
in the Antennae galaxies near this mass (Fritze-v. Alvensleben 1999;
Whitmore et al. 1999).  

The fraction of clusters in the peak can be estimated by computing the
number of initial clusters with masses $700 \le M_{\rm icl}/M_\odot\le
1300$, and comparing this number to the number of all open clusters
that are expected to form in the Milky Way disk in the solar vicinity
and that also enter catalogues, i.e. that have evaporation time-scales
longer than 1~Gyr. These span, roughly, $170 \le M_{\rm
icl}/M_\odot\le 10^4$. The fraction in the peak amounts to 9.1~per
cent for $\sigma_{lM_{\rm ecl}}=0.8$ and to 28~per cent for
$\sigma_{lM_{\rm ecl}}=0.5$.


\section{Constructing an ICMF}
\label{sec:toicmf}
It is interesting that the peak in the ICMF of example~B lies close to
the (classical) initial mass of the Pleiades, Praesepe and Hyades
clusters.  The computations by Portegies Zwart et al. (2001) suggest
that these three clusters form approximately one evolutionary sequence
(\S~\ref{sec:etoi}). This apparent ubiquity of such similar clusters
of different age, and the absence of clusters with initial masses in
the range from a few~$\times10^3$ to a few~$\times10^4\,M_\odot$
within a small region of space around the Sun indicates that the
shoulder seen on Fig.~\ref{fig:mf} around $1000\,M_\odot$ may be a
real feature of the ICMF. The previous section indicates that if
$\sigma_{lM_{\rm ecl}}\approx0.5$ then roughly 30~per cent of all open
clusters may have an initial mass near $1000\,M_\odot$.

To verify the shoulder or otherwise we need an empirical estimate of
the ICMF. To achieve this, the following steps will have to be
undertaken:
\begin{enumerate}
\itemsep=-0.3mm

\item Construct a representative sample of star clusters. A star-burst
galaxy may be used, but here the problem is that the observer does not
know which dynamical state the young clusters are in. A better
approach will be to construct a volume-limited open cluster sample in
the solar vicinity, say out to 500~pc. This has the disadvantage that
the clusters span a very large range of ages and that only the
low-mass part of the CMF is being sampled. The advantage is that the
clusters can be selected to be older than, say, 60~Myr thus ensuring
that the entire evolution driven by gas expulsion is finished
(Fig.~\ref{fig:clev}), and that the Galactic tidal field is reasonably
well known.

\item Construct ``classical'' (\S~\ref{sec:intro}) star-cluster
evolution tracks (e.g. by performing $N-$body computations \'a la
Terlevich 1987; Portegies Zwart et al. 2001).

\item Infer the initial cluster mass of each cluster in the
representative sample by fitting to the classical tracks.

\item Create the ICMF from the ensemble of initial cluster masses.

\end{enumerate}

Any structure within this empirical ICMF should indicate the dominant
physics involved in cluster formation, if it is assumed that the ECMF
is a featureless power-law. Conversely, by applying a favoured
cluster-formation theory (i.e. essentially a model for $f_{\rm
st}(M_{\rm ecl})$) to the so-obtained ICMF, the ECMF may be estimated
for a comparison with the MF of molecular cloud cores. We note in
passing that the MF of star-cluster-forming molecular cloud cores is
related to the ECMF via
\begin{equation}
\xi^{\rm cores}(M_{\rm core}) =\xi^{\rm ecl}(M_{\rm ecl}) \, 
                {dM_{\rm ecl} \over dM_{\rm core}}.
\label{eq:mfcores}
\end{equation} 
If we assume that the SFE does not depend on the mass of the cloud core,
$M_{\rm core} = M_{\rm ecl}/\epsilon$, then
\begin{equation}
\xi^{\rm cores}(M_{\rm core}) = \epsilon\,\xi^{\rm ecl}(M_{\rm ecl}).
\label{eq:mfcores2}
\end{equation} 

\section{The Galactic population~II spheroid}
\label{sec:stsph}
The ancient Galactic stellar spheroid, that consists of population~II
field stars plus globular clusters, totals to a mass of about
$5\times10^7 \le M_{\rm sph}/M_\odot \le 5\times 10^8$ and contains
about 150 globular clusters with masses in the range
$10^4-5\times10^6\,M_\odot$ (e.g. Binney \& Merrifield 1998).  Genesis
of this stellar halo plus globular cluster system is of cosmological
interest, because it pre-dates the formation of the rest of the
Galaxy. The stellar spheroid was born within a time-span of about
1-3~Gyr roughly 13~Gyr ago, corresponding to a SFR of
$0.017-0.5\,M_\odot/{\rm yr}$.

Larsen (2001) notes a well-defined correlation between the absolute
$V$-band magnitude of the brightest cluster, $M_V^{\rm br}$, and the
SFR in the galaxy hosting the star-cluster system. This correlation
implies, for the above SFR, that the most massive cluster formed
during the assembly of the Galactic spheroid may have had $M_V^{\rm
br}\approx-11$ to $-12$ when about 20~Myr old. This corresponds to
globular-cluster masses of roughly $10^{5-6}\,M_\odot$. That
intensively interacting gas-rich galaxies profusely form massive
clusters due to the induced high-pressures in the gas clouds is well
established empirically (Elmegreen et al. 2000), and it is likely that
the Galactic spheroid was likewise assembled during a brief but
violent epoch of merging gas-rich sub-systems.  The question we now
address is if there exists an ECMF that can account for the entire
Galactic spheroid, or if the population~II field stars and the
globular clusters stem from different SF modes.

This problem is approached by assuming the ECMF is a power-law
(eq.~\ref{eq:ecmf}), and that the SF-burst samples clusters with
masses in the range $M_{\rm low}=5\,M_\odot$ to $M_{\rm
max}=1\times10^7\,M_\odot$ from this ECMF. Note that $M_{\rm
max}\approx M_{\omega {\rm Cen}}/ f_{\rm st}$, where $M_{\omega {\rm
Cen}}$ is the mass of the most massive globular cluster, $\omega$~Cen,
and $f_{\rm st}=0.5$ is assumed for this cluster.  The total mass in
the ECMF is then
\begin{equation}
M_{\rm sph} = N_{\rm Gl}\left({1-\beta \over 2-\beta}\right)
              \left( {M_{\rm max}^{2-\beta} - M_{\rm low}^{2-\beta}
              \over   M_{\rm max}^{1-\beta} - M_1^{1-\beta}} 
              \right),
\label{eq:Msph}
\end{equation}
where $N_{\rm Gl}$ is the number of embedded clusters in the
mass-range $M_1=10^4\,M_\odot$ to $M_{\rm max}$. These are the
precursors of the present-day globular clusters.  Vesperini (1998) and
Baumgardt (1998) show that a power-law ECMF with $\beta\approx2$
evolves to the observed Gaussian luminosity function of the
present-day globular cluster system. According to Vesperini's
analysis, $N_{\rm Gl}\approx300$.

\begin{figure}
\begin{center}
\rotatebox{0}{\resizebox{0.6 \textwidth}{!}
{\includegraphics{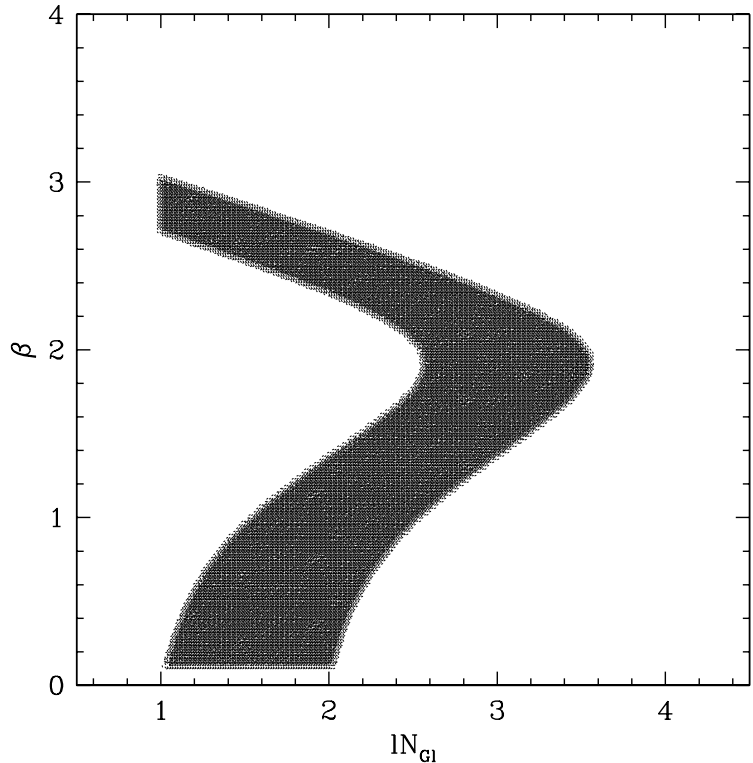}}}
\vskip -20mm
\caption
{The allowed range of ECMFs that can synthesise the population~II
Galactic stellar spheroid. The ECMF is a power law (eq.~\ref{eq:ecmf})
with embedded cluster masses in the mass range $5\le M_{\rm
ecl}/M_\odot \le 1\times 10^7$. The power-law index, $\beta$, is
plotted as a function of the number of clusters, $lN_{\rm Gl}\equiv
{\rm log}_{10}N_{\rm Gl}$, in the mass range $10^4 \le M_{\rm
ecl}/M_\odot \le 1\times 10^7$ needed such that the total stellar
spheroid mass (field stars plus globulars, eq.~\ref{eq:Msph}) is
$5\times 10^7 \le M_{\rm sph}/M_\odot \le 5\times 10^8$.  }
\label{fig:stsph}
\end{center}
\end{figure}

The solution space is shown in Fig.~\ref{fig:stsph}.
The results presented in Fig.~\ref{fig:stsph} suggest that the entire
Galactic spheroid could have been populated by embedded clusters being
sampled from a power-law ECMF with $0.9 < \beta < 2.6$ for $N_{\rm
Gl}\approx 300$. Note that Vesperini's result ($\beta=2, N_{\rm
Gl}=300$) is just compatible with Fig.~\ref{fig:stsph}. It has to be
kept in mind that $N_{\rm Gl}$ is somewhat uncertain given that the
Milky-Way potential and its evolution, which determines how many
clusters are tidally destroyed over a Hubble time, is ill-constrained
(Dehnen \& Binney 1998).

The result obtained here remains valid if the Galactic spheroid 
accumulated through a number of more-or less discrete star-burst
events, as long as each one of them produced a similar ECMF.

\section{Concluding remarks}
\label{sec:conc}
There are reasons to believe that a young star cluster emerges from
the molecular cloud by losing $\simgreat 50$~per cent of its stars.
This fraction, $1-f_{\rm st}$, correlates with the rate with which the
unused gas is expelled. This rate is a function of cluster mass
through the presence or absence of O~stars and the depth of the
potential well of the cluster.  Low-mass clusters that contain no
O~stars (type~I) probably evolve adiabatically, while clusters
containing between about 1000~and $10^5$~stars (type~II embedded
clusters) probably expel their unused gas explosively, that is, more
rapidly than a crossing time.  More massive clusters (type~III) may
again evolve approximately adiabatically because the ionised gas may
not be able to leave the deep potential well within a crossing time.

While the likely behaviour of the function $f_{\rm st}(M_{\rm ecl})$
can be deduced from general physical arguments, important
uncertainties remain.  For example, we do not know yet how good the
approximation $f_{\rm st}\approx0.5$ for adiabatic evolution in a
tidal field is. This needs to be quantified using $N-$body
experiments, and much more work is necessary to quantify the rate with
which gas is expelled from embedded clusters of all mass.

However, the present study demonstrates that a dependence of the
mass-loss rate on embedded cluster mass leads to structure in the ICMF,
even if the ECMF is a featureless power law.  Such structures may be a
gap near $10^4\,M_\odot$ if type~I and type~II embedded clusters
remove their gas faster than a dynamical time.  If, on the other hand,
type~I and type~III clusters rid themselves of natal gas over
time-scales comparable to or longer than a crossing time, while
type~II clusters suffer explosive gas loss, the MF of gas-free but
bound clusters shows a broad depression between about 1000 and
$10^{4.5}\,M_\odot$, while having a pronounced peak near
$10^3\,M_\odot$.

These examples serve to illustrate the type of features that may be
present in the ICMF.  The most important insight gained with this
analysis is that structure in the ICMF may indicate in a statistical
sense the relevant gas-removal processes. It is therefore, at least in
principle, possible to constrain star-cluster formation theories by
studying the ICMF.

An effect not taken into account here but that may affect the ICMF is
the merging of multiple star clusters to form a single more massive
cluster. Also, an ensemble of embedded clusters with equal stellar
mass is likely to have a distribution of SFEs and half-mass radii,
which will broaden features in the resulting ICMF. Such additional
sophistication can be included with a Monte-Carlo approach once
$f_{\rm st}(M_{\rm ecl})$ is better quantified theoretically.

Finally, it was shown that the population~II stellar spheroid plus its
globular clusters can be assembled if all its stars formed in
ensembles of embedded clusters that are sampled from a power-law
ECMF. There is therefore no need to postulate two modes of star
formation, where one mode made the globular clusters and the other
mode the population~II field stars.

\section*{acknowledgements} 

CMB was funded in Heidelberg through the SFB~439 programme until
September 2001. He thanks R.~Spurzem for support.


\vfill 

\end{document}

%% file: mymacros.tex
\newcommand{\getinptfile}
{\typein[\inptfile]{Input file name}
\input{\inptfile}}

\newcommand{\mysummary}[2]{\noi {\bf SUMMARY}#1 \\ \noi \sl #2 \\ \capline 
	\hspace{-.13in} \raisebox{.0in}{$\sqcap$} \rm }  
\newcommand{\mycaption}[2]{\caption[#1]{\footnotesize #2}} 
\newcommand{\capline}{\mbox{}\hrulefill}
\newcommand{\mysection}[2]{ 
\section{\uppercase{\normalsize{\bf #1}}} \def\junksec{{#2}} } %
\newcommand{\mychapter}[2]{ \chapter{#1} \def\junkchap{{#2}}  
\def\thesection{\arabic{chapter}.\arabic{section}}
\def\thesubsection{\thesection.\arabic{subsection}}
\def\thesubsubsection{\thesubsection.\arabic{subsubsection}}
\def\theequation{\arabic{chapter}.\arabic{equation}}
\def\thefigure{\arabic{chapter}.\arabic{figure}}
\def\thetable{\arabic{chapter}.\arabic{table}}
}
\newcommand{\mysubsection}[2]{ \subsection{#1} \def\junksubsec{{#2}} }
\def\thenote{\addtocounter{footnote}{1}$^{\scriptstyle{\arabic{footnote}}}$ }

\newcommand{\myfm}[1]{\mbox{$#1$}}
\def\spose#1{\hbox to 0pt{#1\hss}}	
\def\ltabout{\mathrel{\spose{\lower 3pt\hbox{$\mathchar"218$}} 
     \raise 2.0pt\hbox{$\mathchar"13C$}}}
\def\gtabout{\mathrel{\spose{\lower 3pt\hbox{$\mathchar"218$}}
     \raise 2.0pt\hbox{$\mathchar"13E$}}}
\newcommand{\ltsim}{\raisebox{-0.5ex}{$\;\stackrel{<}{\scriptstyle \backslash}\;$}}
\newcommand{\simlt}{\ltsim}
\newcommand{\simgt}{\gtsim}
%
\newcommand{\unit}[1]{\ifmmode \:\mbox{\rm #1}\else \mbox{#1}\fi}
\newcommand{\ze}{\ifmmode \mbox{z=0}\else \mbox{$z=0$ }\fi }

%
\newcommand{\boldv}[1]{\ifmmode \mbox{\boldmath $ #1$} \else 
 \mbox{\boldmath $#1$} \fi}
%
\renewcommand{\sb}[1]{_{\rm #1}}%
\newcommand{\expec}[1]{\myfm{\left\langle #1 \right\rangle}}
\newcommand{\mone}{\myfm{^{-1}}}
\newcommand{\half}{\myfm{\frac{1}{2}}}
\newcommand{\nth}[1]{\myfm{#1^{\small th}}}
\newcommand{\ten}[1]{\myfm{\times 10^{#1}}}
\newcommand{\abs}[1]{\mid\!\! #1 \!\!\mid}
\newcommand{\as}{a_{\ast}}
\newcommand{\asr}{(a_{\ast}^{2}-R_{\ast}^{2})}
\newcommand{\bvm}{\bv{m}}
\newcommand{\calf}{{\cal F}}
\newcommand{\calI}{{\cal I}}
\newcommand{\calm}{{v/c}}
\newcommand{\calminf}{{(v/c)_{\infty}}}
\newcommand{\calQ}{{\cal Q}}
\newcommand{\calR}{{\cal R}}
\newcommand{\calw}{{\it W}}
\newcommand{\co}{c_{o}}
\newcommand{\cs}{C_{\sigma}}
\newcommand{\cst}{\tilde{C}_{\sigma}}
\newcommand{\cv}{C_{v}}
\def\dbar{{\mathchar '26\mkern-9mud}}	
\newcommand{\deldelr}{\frac{\partial}{\partial r}}
\newcommand{\deldelR}{\frac{\partial}{\partial R}}
\newcommand{\deldeltheta}{\frac{\partial}{\partial \theta} }
\newcommand{\deldelphi}{\frac{\partial}{\partial \phi} }
\newcommand{\ddotrc}{\ddot{R}_{c}}
\newcommand{\ddotxc}{\ddot{x}_{c}}
\newcommand{\dotrc}{\dot{R}_{c}}
\newcommand{\dotxc}{\dot{x}_{c}}
\newcommand{\Estar}{E_{\ast}}
\newcommand{\grpsi}{\Psi_{\ast}^{\prime}}
\newcommand{\kboltz}{k_{\beta}}
\newcommand{\levi}[1]{\epsilon_{#1}}
\newcommand{\limaso}[1]{$#1 ( a_{\ast}\rightarrow 0)\ $}
\newcommand{\limasinfty}[1]{$#1 ( a_{\ast}\rightarrow \infty)\ $}
\newcommand{\limrinfty}[1]{$#1 ( R\rightarrow \infty,t)\ $}
\newcommand{\limro}[1]{$#1 ( R\rightarrow 0,t)\ $}
\newcommand{\limrso}[1]{$#1 (R_{\ast}\rightarrow 0)\ $}
\newcommand{\limxo}[1]{$#1 ( x\rightarrow 0,t)\ $}
\newcommand{\limxso}[1]{$#1 (\xs\rightarrow 0)\ $}
\newcommand{\ls}{l_{\ast}}
\newcommand{\Ls}{L_{\ast}}
\newcommand{\mean}[1]{<#1>}
\newcommand{\ms}{m_{\ast}}
\newcommand{\Ms}{M_{\ast}}
\def\nb{{\sl N}-body }
\def\nbt{{\sf NBODY2} }
\def\nb1{{\sf NBODY1} }
\newcommand{\nuoned}{\nu\sb{1d}}
\newcommand{\ra}{\rightarrow}
\newcommand{\Ra}{\Rightarrow}
\newcommand{\rc}{r_{c} } 
\newcommand{\Rc}{R_{c} } 
\newcommand{\res}[1]{{\rm O}(#1)}
\newcommand{\rnsa}{(r^{2}-a^{2})}
\newcommand{\Rnsa}{(R^{2}-a^{2})}
\newcommand{\rs}{r_{\ast}}
\newcommand{\Rs}{R_{\ast}}
\newcommand{\Rsa}{(R_{\ast}^{2}-a_{\ast}^{2})}
\newcommand{\sa}{\sigma } 
\newcommand{\sac}{\sigma_{c} } 
\newcommand{\sas}{\sigma_{\ast} } 
\newcommand{\sasp}{\sigma^{\prime}_{\ast}}
\newcommand{\saxs}{\sigma_{\ast} } 
\newcommand{\sech}{{\rm sech}}
\newcommand{\tff}{t\sb{ff}} 
\newcommand{\ti}{\tilde}
\newcommand{\trel}{t\sb{rel}}
\newcommand{\ts}{\tilde{\sigma} } 
\newcommand{\tss}{\tilde{\sigma}_{\ast} } 
\newcommand{\vcol}{v\sb{col}}
\newcommand{\vs}{v_{\ast}  } 
\newcommand{\vsp}{v^{\prime}_{\ast}}
\newcommand{\vxs}{v_{\ast}  } 
\newcommand{\xs}{x_{\ast}}
\newcommand{\xc}{x_{c} } 
\newcommand{\xistar}{\xi_{\ast}}
\newcommand{\rmd}{\ifmmode \:\mbox{{\rm d}}\else \mbox{ d}\fi }
\newcommand{\rmD}{\ifmmode \:\mbox{{\rm D}}\else \mbox{ D}\fi }
\newcommand{\valfven}{v_{{\rm Alfv\acute{e}n}}}

%
\newcommand{\noi}{\noindent}
\newcommand{\bc}{boundary condition }
\newcommand{\bcs}{boundary conditions }
\newcommand{\Bcs}{Boundary conditions }
\newcommand{\lhs}{left-hand side }
\newcommand{\rhs}{right-hand side }
\newcommand{\wrt}{with respect to }
\newcommand{\iras}{{\sl IRAS }}
\newcommand{\cobe}{{\sl COBE }}
\newcommand{\Oh}{\myfm{\Omega h}}
%
\newcommand{\etal}{{\em et al.\/ }}
\newcommand{\eg}{{\em e.g.\/ }}
\newcommand{\etc}{{\em etc.\/ }}
\newcommand{\ie}{{\em i.e.\/ }}
\newcommand{\viz}{{\em viz.\/ }}
\newcommand{\cf}{{\em cf.\/ }}
\newcommand{\via}{{\em via\/ }}
\newcommand{\apriori}{{\em a priori\/ }}
\newcommand{\adhoc}{{\em ad hoc\/ }}
\newcommand{\viceversa}{{\em vice versa\/ }}
\newcommand{\versus}{{\em versus\/ }}
\newcommand{\qed}{{\em q.e.d. \/}}
\newcommand{\<}{\thinspace}
%
\newcommand{\km}{\unit{km}}
\newcommand{\kms}{\unit{km~s\mone}}
\newcommand{\kmsa}{\unit{km~s\mone~arcmin}}
\newcommand{\kpc}{\unit{kpc}}
\newcommand{\mpc}{\unit{Mpc}}
\newcommand{\hkpc}{\myfm{h\mone}\kpc}
\newcommand{\hmpc}{\myfm{h\mone}\mpc}
\newcommand{\parsec}{\unit{pc}}
\newcommand{\cm}{\unit{cm}}
\newcommand{\yr}{\unit{yr}}
\newcommand{\au}{\unit{A.U.}}
\newcommand{\AU}{\au}
\newcommand{\gm}{\unit{g}}
\newcommand{\solarm}{\unit{M\sun}}
\newcommand{\Lsun}{\unit{L\sun}}
\newcommand{\Rsun}{\unit{R\sun}}
\newcommand{\seconds}{\unit{s}}
\newcommand{\micro}{\myfm{\mu}}
\newcommand{\Mdot}{\myfm{\dot M}}
%
%
%
\newcommand{\dgr}{\myfm{^\circ} }
\newcommand{\ddgr}{\mbox{\dgr\hskip-0.3em .}}
\newcommand{\mnt}{\mbox{\myfm{'}\hskip-0.3em .}}
\newcommand{\scnd}{\mbox{\myfm{''}\hskip-0.3em .}}
\newcommand{\hr}{\myfm{^{\rm h}}}
\newcommand{\dhr}{\mbox{\hr\hskip-0.3em .}}
%
%
%
%
%
%
%
\newcommand{\refindent}{\par\noindent\hangindent=0.5in\hangafter=1}
\newcommand{\figpar}{\par\noindent\hangindent=0.7in\hangafter=1}
%
%

\newcommand{\mybiblio}{\vspace{1cm}
		       \setcounter{subsection}{0}
		       \addtocounter{section}{1}
		       \def\junksec{References} 
 }

%
%
%

%
%
%
%
%

\newcommand{\vol}[2]{ {\bf#1}, #2}
\newcommand{\jour}[4]{#1. {\it #2\/}, {\bf#3}, #4}
\newcommand{\physrevd}[3]{\jour{#1}{Phys Rev D}{#2}{#3}}
\newcommand{\physrevlett}[3]{\jour{#1}{Phys Rev Lett}{#2}{#3}}
\newcommand{\aaa}[3]{\jour{#1}{A\&A}{#2}{#3}}
\newcommand{\aaarev}[3]{\jour{#1}{A\&A Review}{#2}{#3}}
\newcommand{\aaas}[3]{\jour{#1}{A\&A Supp.}{#2}{#3}}
\newcommand{\aj}[3]{\jour{#1}{AJ}{#2}{#3}}
\newcommand{\apj}[3]{\jour{#1}{ApJ}{#2}{#3}}
\newcommand{\apjl}[3]{\jour{#1}{ApJ Lett.}{#2}{#3}}
\newcommand{\apjs}[3]{\jour{#1}{ApJ Suppl.}{#2}{#3}}
\newcommand{\araa}[3]{\jour{#1}{ARAA}{#2}{#3}}
\newcommand{\mn}[3]{\jour{#1}{MNRAS}{#2}{#3}}
\newcommand{\mnras}{\mn}
\newcommand{\jgeo}[3]{\jour{#1}{Journal of Geophysical Research}{#2}{#3}}
\newcommand{\qjras}[3]{\jour{#1}{QJRAS}{#2}{#3}}
\newcommand{\nat}[3]{\jour{#1}{Nature}{#2}{#3}}
\newcommand{\pasa}[3]{\jour{#1}{PAS Australia}{#2}{#3}}
\newcommand{\pasj}[3]{\jour{#1}{PAS Japan}{#2}{#3}}
\newcommand{\pasp}[3]{\jour{#1}{PAS Pacific}{#2}{#3}}
\newcommand{\rmp}[3]{\jour{#1}{Rev. Mod. Phys.}{#2}{#3}}
\newcommand{\science}[3]{\jour{#1}{Science}{#2}{#3}}
\newcommand{\vistas}[3]{\jour{#1}{Vistas in Astronomy}{#2}{#3}}